\documentclass{ws-ijmpd}

\usepackage{subfigure}
\usepackage{epsfig}
\usepackage{psfig}
\usepackage{amsfonts}
\usepackage{amsmath}
\usepackage{amssymb}
\usepackage{times}
\begin{document}

\markboth{Zhang \& Qin}
{Parameters of the prompt GRB emission}

\catchline{}{}{}{}{}
%
%%%%%%%%%%%%%%%%%%%%%%%%%%%%%%%%%%%%%%%%%%%%%%%%%%%%%%%%%%%%%%%%%%%%

\title{Parameters of the prompt gamma-ray burst
emission estimated with the opening angle of jets\footnote{send
offprint requests to: Y.-P. Qin: ypqin@ynao.ac.cn}  }

\author{B.-B. Zhang\footnote{E-mail:zbinbin@ynao.ac.cn}}

\address{National Astronomical Observatories/Yunnan
Observatory, Chinese Academy of Sciences\\
The Graduate School of the Chinese Academy of Sciences \\
P.O.Box 110, Kunming,Yunnan, 650011, P. R. China\\}

\author{Y.-P. Qin}

\address{National Astronomical Observatories/Yunnan
Observatory, Chinese Academy of Sciences,\\
P. O. Box 110, Kunming,Yunnan, 650011, P. R. China\\
Physics Department, Guangxi University, Nanning, Guangxi
530004, P. R. China}

\maketitle

\begin{history}
\received{Day Month Year}
\revised{Day Month Year}
\comby{Managing Editor}
\end{history}

\begin{abstract}
We present in this paper an approach to estimate the initial
Lorentz factor of gamma-ray bursts (GRBs) without referring to the
delayed emission of the early afterglow. Under the assumption that
the afterglow of the bursts concerned occurs well before the
prompt emission dies away, the Lorentz factor measured at the time
when the duration of the prompt emission is ended could be
estimated by applying the well-known relations of GRB jets. With
the concept of the efficiency for converting the explosion energy
to radiation, this Lorentz factor can be related to the initial
Lorentz factor of the source. The corresponding rest frame peak
energy can accordingly be calculated. Applying this method, we
estimate the initial Lorentz factor of the bulk motion and the
corresponding rest frame spectral peak energy of GRBs for a new
sample where the redshift and the break time in the afterglow are
known. Our analysis shows that, in the circumstances, the initial
Lorentz factor of the sample would peak at $200$ and would be
distributed mainly within $(100,400)$, and the peak of the
distribution of the corresponding rest frame peak energy would be 
$0.8keV$ and its main region would be $(0.3keV,3keV)$.
\end{abstract}


\section{Introduction}	

One of the recent exiting discoveries in gamma-ray bursts (GRBs)
is the break detected in the afterglow light curve of some bursts
which could be interpreted as a consequence of the beamed emission
(e.g., Ref. \refcite{Rho97},\refcite{sar99}). If the afterglow emission
is beamed, how is the prompt emission? Whether the latter emission
is isotropic or strongly beamed in our direction has been an open
question for some years. As mentioned in Ref. \refcite{sar99}, this
question has implications on almost every aspect of the
phenomenon, from the energetics of the events to the engineering
of the inner engine and the statistics and the luminosity function
of the sources. Frail et al. (2001) studied\cite{fra01} a sample of GRBs with
good afterglow follow-up and known redshifts. They interpreted the
breaks in the scenario of the beamed model and found that most
bursts with large values of the isotropic-equivalent gamma-ray
energy, $E_{iso}$, possess the smallest beaming fraction,
$f_b=1-\cos \theta _{jet}$. The collimation-corrected energy,
$E_\gamma =f_bE_{iso}$, of this sample is strongly clustered. This
was independently confirmed by Panaitescu \& Kumar (2001)\cite{pan01a}. Bloom
et al.(2003) collected a larger sample of GRBs and found that the distribution of $%
E_\gamma $ clusters around $1.3\times 10^{51}$ ergs\cite{blo03}. All these
were regarded as evidence supporting the beamed emission scenario.

The isotropic gamma-ray energy $E_{iso}$ was found to be
correlated with the cosmological co-moving frame peak energy
$E_{p}$ by different authors (see, e.g., Refs.~\refcite{llo00}--\refcite{yon04}). More recently, Ghirlanda et al. (2004) found\cite{ghi04} a tight
correlation between $E_{\gamma }$ and $E_{p}$, which sheds light
on the still uncertain radiation processes for the prompt GRB
emission. In computing $E_{\gamma }$, it is essential that the
beaming fraction is available. According to Refs.~\refcite{sar99},
the opening angle of the jet can be calculated with
\begin{equation}
\theta _{jet}=B(\frac{t_{jet}}{1+z})^{3/8}(\frac{\xi
n}{E_{iso}})^{1/8}
\end{equation}%
in the case of a homogeneous circumburst medium, where $B$ is a
constant which can be found in Ref.~\refcite{fri05}, $t_{jet}$
is the afterglow jet break time, $z$ is the redshift, $\xi $ is
the efficiency for converting the explosion energy to radiation,
$n$ is the density of the ambient medium, and $E_{iso}$ is the
energy in $\gamma $-rays calculated assuming that the emission is
isotropic. This enables us to estimate the opening angle of jets
and with it to peep into other parameters associated with the
mechanism of radiation.

Indeed, under the scenario of jets, parameters such as the total
energy in the relativistic ejecta, the jet opening angle, the
density and profile of the medium in the immediate vicinity of the
burst, and those associated with the microphysical shocks could be
estimated by modeling the broadband emission of GRB afterglows for
various bursts (see Refs.~\refcite{pan01a},\refcite{pan01}--\refcite{pan02}). Assuming that the observed GRB
durations are a good measure of the ejecta deceleration timescale,
the jet Lorentz factors at the deceleration radius were found to
be within 70 and 300 for 10 GRBs\cite{pan02}.

In addition to the break time observed in the afterglow, Sari \&
Piran (1999) suggested\cite{sar99a} that the reverse shock of a burst could
provide a crude measurement of the initial Lorentz factor. In the
case of GRB 990123, they showed that the initial Lorentz factor
$\Gamma \sim 200$ could be obtained from the prompt optical flash
observed in the burst.

Early afterglow which could overlap the main burst was predicted
previously (see, e.g., Ref.~\refcite{sar97}). According to the analysis of
Ref. ~\refcite{sar99b} based on the internal-external shocks model,
for short bursts the peak of the afterglow will be delayed,
typically, by few dozens of seconds after the burst while for long
ones the early afterglow emission will overlap the GRB signal, and
a delayed emission, with the characteristics of the early
afterglow, can be used to measure the initial Lorentz factor of
the relativistic flow (see, also Refs.~\refcite{sar99c}--\refcite{nak05}).

Hinted by the previous works, we wonder if the initial Lorentz
factor could be estimated with an independent approach in case the
delayed emission is not available. An investigation on this issue
is organized as follows. In section 2, we present appropriate
formulas and show how to estimate the concerned parameters with
them. A new GRB sample for which the redshift and the break time
in the afterglow are known is studied in section 3. Conclusions
are summarized in the last section, where a brief discussion is
present.

\section{Formulas employed}

As the initial explosion of GRBs is not at all in the stage of
afterglow, we cannot estimate parameters of the former from the
measurements of the latter according to the law of the afterglow.
To connect the two phases, we need to find a particular moment
which satisfies the following requirements: a) at that moment, the
afterglow has already begun so that the law of the afterglow is
applicable; b) parameters associated with that moment are well
related with those of the initial explosion.

As predicted above, for long bursts the early afterglow emission
will overlap the GRB signal. Indeed, it was reported recently that
a bright optical emission from GRB 990123 was detected while the
burst was still in progress\cite{ake99}. Revealed in Ref.~\refcite{fox03}
is the discovery of the optical counterpart of GRB
021004 only 193 seconds after the event. The time (measured from
the trigger) is slightly longer than the duration of the event. Li
et al. (2003) showed\cite{li03} that the faintness of GRB 021211, coupled
with the fast decline typical of optical afterglows, suggests that
some of the dark bursts were not detected because optical
observations commenced too late after the GRB.

Accordingly, we assume that to the time the prompt emission of
GRBs is going to be undetectable, which is generally represented
by the concept of duration $t_{dur}$ (in BATSE, it is associated
with $T_{90}$), the afterglow of the source has already emerged.
Under this assumption, $t_{dur}$ is the moment that can satisfy
the first requirement. Shown in the following one will find that
$t_{dur}$ can also satisfy the second requirement. In addition,
$t_{dur}$ is fortunately always available.

The observed complex structure of GRB\ light curves suggests that,
during the prompt emission of a burst, several ejecta with
different masses and different Lorentz factors might be involved.
We assume that, after the process of the prompt emission, all
ejecta are merged as a single one which has mass $m$ and bears a
Lorentz factor $\Gamma _{dur}$ (which is measured at the end of
the duration of the burst). Let's define an average initial
Lorentz factor of the early phase ejecta as%
\begin{equation}
\Gamma \equiv \frac{1}{m}\sum_{i}\Gamma _{i}m_{i},
\end{equation}%
with%
\begin{equation}
m=\sum_{i}m_{i},
\end{equation}%
where $\Gamma _{i}$ and $m_{i}$ are the initial Lorentz factor and
the mass of the $i$th ejecta, respectively. One can check that,
for a burst associated with a shock produced by the collision of
two shells with roughly equal masses, $\Gamma \simeq (\Gamma
_{in}+\Gamma _{out})/2$, where $\Gamma _{in}$ and $\Gamma _{out}$
are the Lorentz factors of the inner and outer shells,
respectively, while for a burst containing several shells with
roughly equal masses, $\Gamma \simeq \sum_{i}\Gamma _{i}/N$, where
$N$ is the number of shells. Note that the initial kinetic energy
of all these early phase ejecta is the product of the explosion of
the burst (where a number of sub-explosions might be involved). It
is well known that it is the losing of the kinetic energy of the
GRB ejecta that gives rise to the energy of the radiation observed
during the prompt emission as well as the increasing of the
thermal energy during this period, regardless what the radiation
mechanism is. According to the conservation of energy, one finds
\begin{equation}
\Gamma mc^{2}-\Gamma _{dur}mc^{2} =\xi (\Gamma -1)mc^{2}+\Delta
E_{th},
\end{equation}%
where $\Delta E_{th}$ is the increasing of the thermal energy of
the system. Assuming that the radiation is associated with the
synchrotron mechanism, as is generally believed, then parts of the
increasing thermal energy at any moment must be converted to
radiation due to the increasing velocity of individual electrons.
Based on this argument, we believe that, as a sum of that of all
moments, $\Delta E_{th}$ would be significantly reduced, compared
with that obtained in the situation where the synchrotron
mechanism is not at work. We accordingly assume that, during the
shock, the increasing thermal energy is much smaller than the
radiation energy. That is, we assume $\xi (\Gamma
-1)mc^{2}\gg\Delta E_{th}$. Omitting the increasing of the thermal
energy, we get from (4) that
\begin{equation}
(1-\xi )\Gamma  \simeq\Gamma _{dur}-\xi .
\end{equation}%
When all the initial kinetic energy is converted to photons, we
have $\xi =1$ and then $\Gamma _{dur}=1$, and when none of the
initial kinetic energy is changed to radiation, we get $\xi =0$
and then $\Gamma _{dur}=\Gamma $. Thus, one always finds $1\leq
\Gamma _{dur}\leq \Gamma $.
According to (5), the initial Lorentz factor could be determined as long as $%
\Gamma _{dur}$ and $\xi $ (when $\xi \neq 1$) are known.

During the period of the afterglow, when the external matter is
homogenously distributed, the Lorentz factor would decline
following the law of $\Gamma (t)\propto t^{-p}$, where $p=3/7$ in
a radiative phase and $p=3/8$ in an adiabatic phase (see, e.g.,
Ref.~\refcite{pir05}). Thus, we get $\Gamma
_{dur}=(t_{jet}/t_{dur})^{p}\Gamma _{jet}$, where $\Gamma _{jet}$
is the Lorentz factor of the ejecta measured at $t_{jet}$.
According to the beamed model, a break in the afterglow light
curve of the burst would appear when
its bulk Lorentz factor becomes of the order of $1/\theta _{jet}$, i.e., $%
\Gamma _{jet}\simeq 1/\theta _{jet}$. We then come to
\begin{equation}
\Gamma _{dur}\simeq (\frac{t_{jet}}{t_{dur}})^{p}\frac{1}{\theta
_{jet}}.
\end{equation}%
For $\xi <1$, we get from equations (5) and (6) that
\begin{equation}
\Gamma \simeq \frac{(t_{jet}/t_{dur})^{p}/\theta _{jet}-1}{1-\xi
}+1.
\end{equation}

As is generally assumed, the jet of bursts is strongly beamed in
our direction so that the emission is detectable due to the great
Doppler boosting (see, e.g., Ref.~\refcite{sar99}). According to the
Doppler effect, a photon of $E_{0}$ emitted from the area of
$\theta =0$ within the spherical
surface of a uniform jet which moves outwards with a bulk Lorentz factor $%
\Gamma $ would be blue-shifted to $E=2\Gamma E_{0}$. In the case
of photons being emitted from a certain area with a rest frame
Band function spectrum\cite{ban93} which peaks at
$E_{0,p}$, the spectrum would be blue-shifted and would peak at
$E_{p}$ which is proportional to $E_{0,p}$ (see Table 4 in Ref.~\refcite{qin02}
where $E_{p}=1.67\Gamma E_{0,p}$ can be concluded). Neglecting the
minute difference we take in the following that $E_{p}\simeq
2\Gamma E_{0,p}$. Following Ref.~\refcite{sar99}, we consider
through out this paper only an adiabatic phase and then take
$p=3/8$. Thus, from equation (7) we get
\begin{equation}
\Gamma \simeq \frac{(t_{jet}/t_{dur})^{3/8}/\theta _{jet}-1}{1-\xi
}+1
\end{equation}%
and
\begin{equation}
E_{0,p}\simeq \frac{(1-\xi
)E_{p}/2}{(t_{jet}/t_{dur})^{3/8}/\theta _{jet}-\xi }.
\end{equation}

\section{Application}

Presented in Ref.~\refcite{fri05} are 52 GRB or XRF sources
(called the FB sample) where their redshifts as well as the
gamma-ray fluences are available. The isotropic energies $E_{iso}$
were calculated assuming a standard cosmology of $(\Omega
_{M},\Omega _{\Lambda },h)=(0.3,0.7,0.7)$. For some of these
sources, break times $t_{jet}$ are available, and then with
equation (1) the opening angles $\theta _{jet}$ of the sources
could be well determined, where $\xi =0.2$ is assumed (see also
Ref.~\refcite{fra01}). According to equations (8) and (9), to
calculate $\Gamma $ and $E_{0,p}$ for these sources we need to
know $t_{dur}$ as well. Listed in Table 1 are the values of
$t_{dur}$, which is measured in various bands, for the sources of
the FB sample with $t_{jet}$, $\theta _{jet}$, and $E_{p}$
available. To
meet the requirement that the afterglow has already begun (i.e., $%
t_{aft}\leq t_{dur}$, where $t_{aft}$ is the start time of the
afterglow) and the prompt emission is just ended so that the
common value of the efficiency $\xi $ can be adopted and the mass
of the piled up ambient medium is relatively small, we adopt the
largest value of $t_{dur}$ to calculate $\Gamma $ and $E_{0,p}$.
The results are presented in Table 2.
\newpage
\begin{table}[ph]
% \centering
\tbl{Data of $t_{dur}$}
 { \begin{tabular*}{8cm}{@{}ccccl@{}} \toprule
GRB & trig. NO. & Duration Band & $t_{dur}$ &Ref. \\
(/XRF) &   & ($Kev$) & (s) &  \\
\hline %dddddddddd
970508 & 6225  & $20 \sim 1000$ & 35 & \refcite{kou97}\\
 &  &$50\sim300$ & 35 & \refcite{kou97} \\
 &  & -- & 15 & \refcite{cos97} \\
 &  &$25\sim(>320)$&23.104(3.789) & BATSE \\
970828 & 6350   & $2\sim12$ & 160 & \refcite{rem98} \\
980519 & 6764  & $40\sim700$ & 30 & \refcite{mul98}\\
  &  & $50\sim300$ & 60 & \refcite{con98}\\
  &  & $2\sim28$ & 190 &  \refcite{mul98} \\
 &  &$25\sim(>320)$&23.808(1.032) &BATSE \\
 980703 & 6891  & $40\sim700$ & 90 & \refcite{ama98} \\
  &  & $50\sim300$ & 400 & \refcite{kip98}\\
  &  & $2\sim12$ & 40 & \refcite{smi98} \\
  &  & $2\sim20$ & 400 & \refcite{kip98} \\
 &  &$25\sim(>320)$&411.648(9.273) &BATSE \\
990123 & 7343  & $40\sim700$ & 100 & \refcite{fer99}  \\
  &  & $50\sim300$ & 63.3 & \refcite{kip99a} \\
 &  & $20\sim1000$ & 63.3 & \refcite{kip99a} \\
 &  &$25\sim(>320)$&63.36(0.264 ) & BATSE\\
 990510 & 7560 & $40\sim700$ & 80 & \refcite{dad99} \\
  &  & $50\sim300$ & 100 & \refcite{kip99b}\\
  &  & $20\sim100$ & 100 & \refcite{kip99b} \\
 &  &$25\sim(>320)$&68.032(0.202) & BATSE\\
 990705 & 7633 &   $40\sim700$ & 45 & \refcite{cel99} \\
   &  & $2\sim26$ & 45 & \refcite{cel99} \\
 990712 & 7648  & $40\sim700$ & 30 & \refcite{hei99} \\
 &  &$25\sim(>320)$&31.616(3.137) & BATSE\\
 991216 & 7906 &  $50\sim300$ & 50 & \refcite{kip99} \\
 &    & $20\sim100$ & 50 & \refcite{kip99} \\
 &  &$25\sim(>320)$&15.168(0.091) & BATSE\\
011211   &  & $40\sim700$ & 270 & \refcite{fro02}\\
020124   &  & $8\sim85$ & 70 & \refcite{ric02}\\
020405   &  & $25\sim100$ & 40 & \refcite{hur02a} \\
020813   &  & $2\sim25$ & 125 &  \refcite{vil02} \\
 &  &   $25\sim100$ & 125 & \refcite{hur02b} \\
 &  &   $8\sim40$ & 125 & \refcite{hur02b} \\
021004   &  & $8\sim40$ & 100 & \refcite{shi02} \\
021211   &  & $8\sim40$ & 5.7 & \refcite{cre02} \\
030226   &  & $30\sim400$ & 100 & \refcite{suz03} \\
030328   &  & $30\sim400$ & 100 & \refcite{vil03} \\
030329   &  & $30\sim400$ & 50 & \refcite{ric03}\\
 &  &   $15\sim5000$ & 35 & \refcite{gol03}\\
030429   &  & $30\sim400$ & 14 & \refcite{dot03} \\
 &  &   $25\sim100$ & 5 & \refcite{hur03} \\
040511   &  & $30\sim400$ & 38 & \refcite{dul04} \\
041006  & & $25\sim100$ & 24.6 & \refcite{van04} \\
\hline
\end{tabular*} \label{ta1}}
\end{table}
\newpage
\begin{table}[ph]
% \begin{minipage}{70mm}
  \tbl{Estimated values of the initial Lorentz factor and the corresponding rest
frame peak energy}
 {  \begin{tabular}{@{}ccccc@{}}
  \\
   \hline
 GRB/XRF & $E_{0,p}$ & $\Gamma$\\
\hline
970508&  0.354(0.111) &  206(26) \\
970828&  2.081(0.477) &  141(16) \\
980519&  2.311(0.594) &  157(26) \\
980703&  3.397(0.734) &   75( 7) \\
990123&  4.027(0.630) &  253(34) \\
990510&  0.752(0.086) &  285(17) \\
990705&  0.794(0.116) &  220(27) \\
990712&  0.330(0.065) &  142(14) \\
991216&  1.193(0.335) &  270(54) \\
011211&  0.753(0.101) &  124( 8) \\
020124&  0.723(0.183) &  254(48) \\
020405&  1.530(0.405) &  202(36) \\
020813&  0.857(0.114) &  188(18) \\
021004&  0.930(0.521) &  144(51) \\
021211&  0.133(0.033) &  343(68) \\
030226&  0.859(0.203) &  170(15) \\
030328&  0.837(0.117) &  191(18) \\
030329&  0.293(0.024) &  136(10) \\
030429&  0.203(0.083) &  317(97) \\
040511&  0.697(0.189) &  236(44) \\
041006&  0.289(0.106) &  190(59) \\
\hline
\end{tabular}}
%\end{minipage}
\end{table}

Shown in Fig. 1 is the relation between $E_{0,p}$ and $\Gamma$. It
shows clearly that $\log E_{0,p}$ and $\log \Gamma $ are not
correlated at all. The un-correlation between the two quantities
seems to suggest that the rest frame peak energy is strongly
associated with the mechanism rather than with the expansion speed
(one will find in the following that the distribution of $\Gamma $
is much more clustered than that of $E_{0,p}$).
%%%%%%%%%%%%%%5fig1
\begin{figure}
\centering
\includegraphics[width=3.0in]{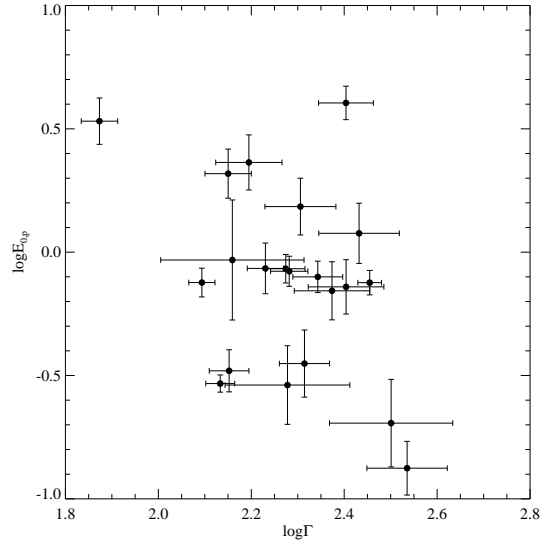}
\caption{Relationship between $E_{0,p}$ and $\Gamma $. The
correlation coefficient between $\log E_{0,p}$ and $\log \Gamma $ is $%
r=0.078$ ($N=21$) and the probability of rejecting the null hypothesis is 0.734}
\end{figure}

Displayed in Fig. 2 are the distributions of $E_{0,p}$ and
$\Gamma$. The rest frame peak energy peaks at $E_{0,p}=0.8keV$ and is mainly distributed
within $(0.3keV,3keV)$. The Lorentz factor peaks at $200$ and it
is found mainly within $(100,400)$, which is very narrow.
\begin{figure}
  \centering
  \subfigure[]{
    \label{fig:subfig:a} %% label for first subfigure
    \includegraphics[width=3.0in]{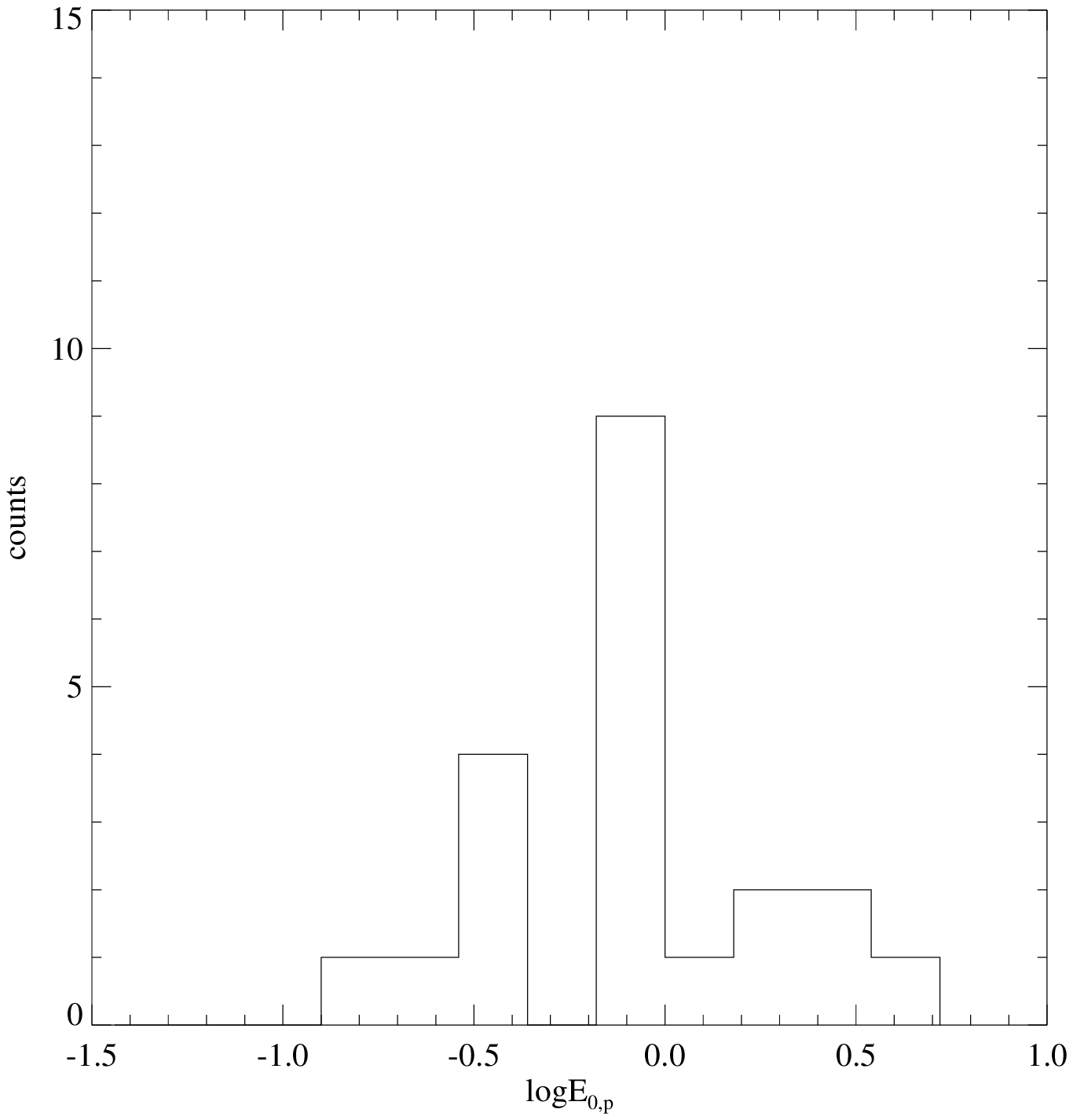}}
 % \hspace{1in}
  \subfigure[]{
    \label{fig:subfig:b} %% label for second subfigure
    \includegraphics[width=3.0in]{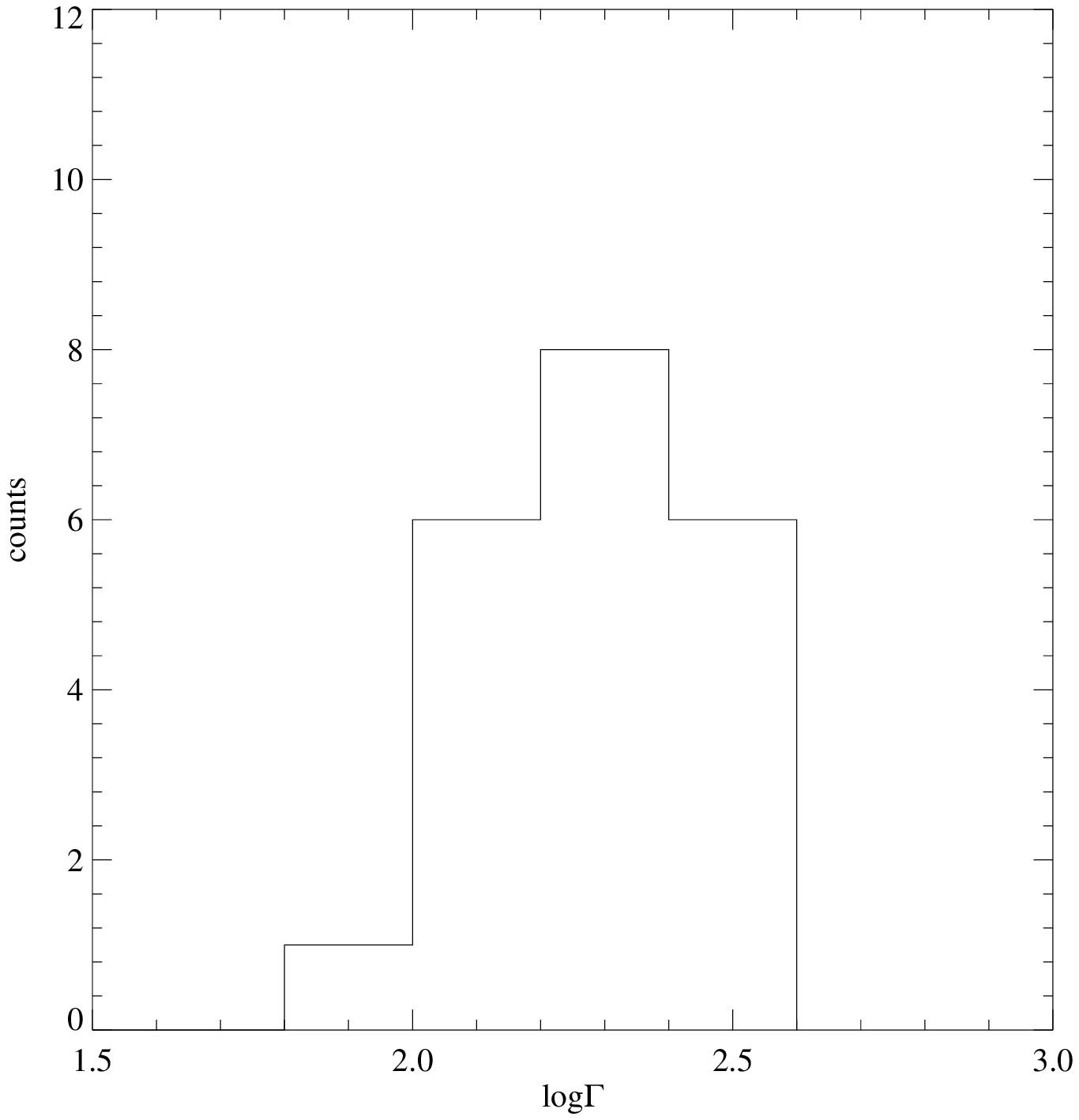}}
 \caption{ Distributions (solid lines) of $\log
E_{0,p}$ (a) and $\log \Gamma $ (b).}
\end{figure}

As Fig. 3 shows, there is a very tight correlation between $E_{0,p}$ and $%
E_p $: $\log E_p=(0.85\pm 0.25)\log E_{0,p}+(2.56\pm 0.10)$. Note
that, relation $E_p\simeq 2\Gamma E_{0,p}$ itself could not
guarantee the
correlation, since if it did, it should also lead to a correlation between $%
E_{0,p}$ and $\Gamma $, but this is not true (see Fig. 1). The
correlation between $E_{0,p} $ and $E_p$ must arise from
mechanisms other than from the Doppler effect.
\begin{figure}
\centering
\includegraphics[width=3.0in]{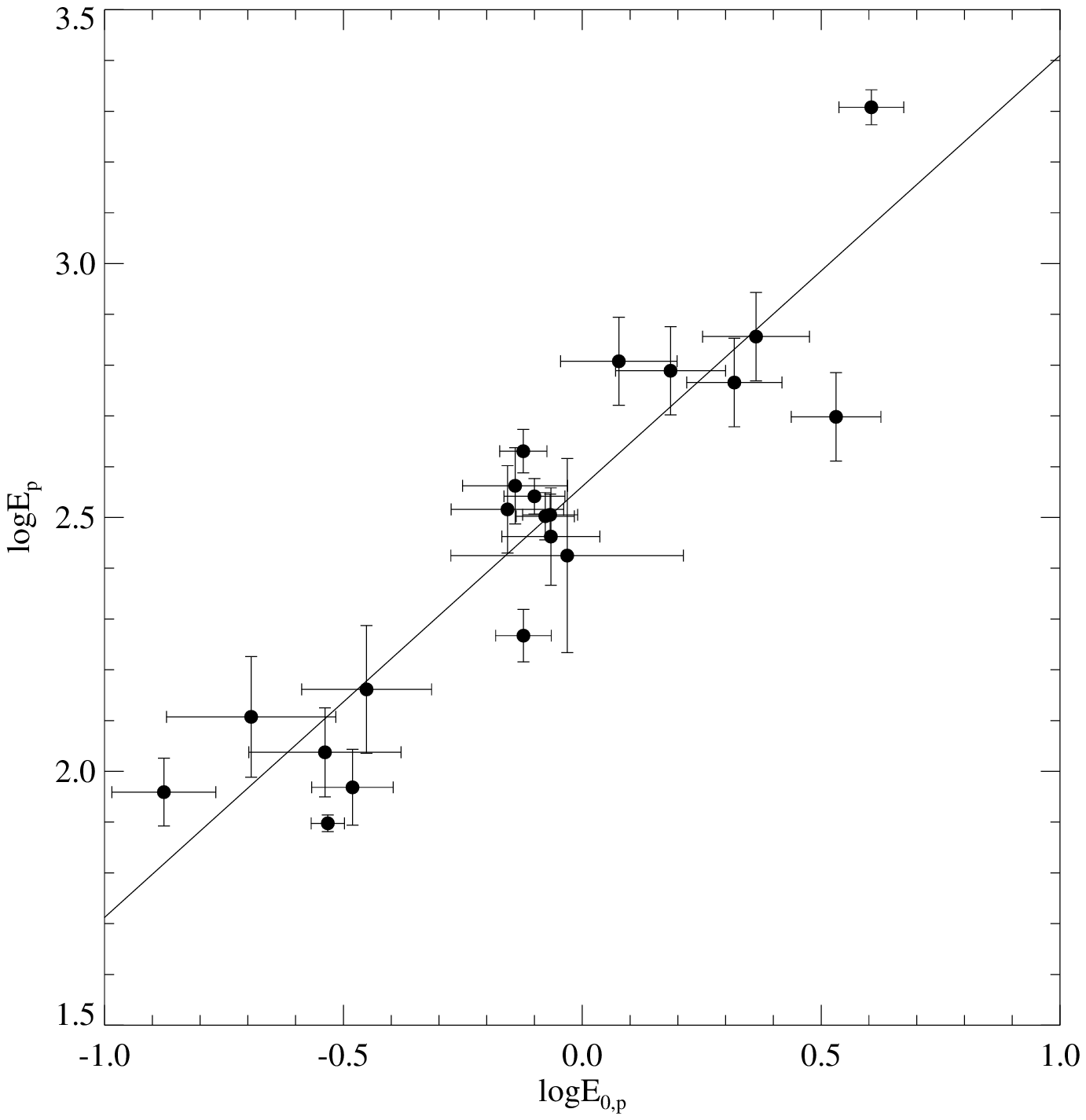}
\caption{Relationship between $E_p$ and $E_{0,p}$. The correlation
coefficient between $\log E_p$ and $\log E_{0,p}$ is $r=0.960$
($N=21$) and the probability of rejecting the null hypothesis is
$P<0.0001$.}
\end{figure}
\section{Discussion and conclusions}

In this paper we propose a method which does not refer to the
delayed emission of the early afterglow to estimate the initial
Lorentz factor of GRBs, in case the detection of the early
afterglows of many bursts might be missed. Due to the fact that the
afterglows of some bursts were observed soon after the detection of
the main emission, we assume that the afterglows of the bursts
concerned occur well before the prompt emission dies away (i.e.,
we assume $t_{aft}\leq t_{dur}$). Under this assumption, the bulk
Lorentz factor of a burst measured at the break time, $t_{jet}$,
and that measured at the time marking the end of duration,
$t_{dur}$, could be well related by the law of $\Gamma (t)\propto
t^{-3/8}$ according to the beaming scenario. Employing the concept
of the efficiency for converting the explosion energy to
radiation, $\xi $, we can relate the initial Lorentz factor of a
burst to that measured at $t_{dur}$. Combining the two relations
one can therefore estimate the initial Lorentz factor of a burst
from that measured at $t_{jet}$. The corresponding rest frame peak
energy can hence be estimated from this initial Lorentz factor and
the observed peak energy according to the Doppler effect.

Applying this method, the initial Lorentz factors of the bulk
motion as well as the corresponding rest frame spectral peak
energies of GRBs for a new sample for which the redshift and the
break time in the afterglows are available are estimated. The
sample employed is that presented currently by Ref.~\refcite{fri05}.
 Our analysis shows that the initial Lorentz factor $\Gamma
$ peaks at $200$ and is distributed mainly within $(100,400)$, and
the peak of the distribution of the corresponding rest frame peak energy is $%
E_{0,p}=0.8keV$ and its main region is $(0.3keV,3keV)$. 

It is known that, a large value of the Lorentz factor, $\Gamma
>100$, is essential to overcome the compactness problem (see,
e.g., Ref.~\refcite{pir05}). As individual cases, the optical flash
accompanying GRB 990123 provids a direct evidence for a large
Lorentz factor\cite{sar99a} $\Gamma \sim 200$ 
Statistically, Mallozzi et al. (1995) found\cite{mal95} that the average value
of $E_p$ for 82 bright bursts is $\sim 340keV$. Taking
$E_{0,p}=0.8keV$ and adopting $E_p\simeq 2\Gamma E_{0,p}$, we find
that the average Lorentz factor of these bursts would be $\sim
213$, which is consistent with what we obtained above. Preece et
al. (2000) revealed\cite{pre00} by the analysis of high time resolution
spectroscopy of 156 bright bursts that the main range of $E_p$ for
these sources could be found to be within $\sim [100,800]keV$.
This would lead to a range of $\Gamma \sim [62,500]$ when adopting
$E_{0,p}=0.8keV$ and $E_p\simeq 2\Gamma E_{0,p}$, which is also in
agreement with what we find in this paper.

As shown in Table 2, the estimated initial Lorentz factor for GRB 990123 is $%
\Gamma \sim 253\pm34 $ which is slightly larger than what is obtained with
the method referring to the delayed emission of the early
afterglow (see Ref.~\refcite{sar99a}, where $\Gamma \sim 200$ was
presented). Applying (6), we get $\Gamma _{dur}=202\pm22$ for this
source. We argue that the initial Lorentz factor estimated with
our method is that associated with the initial explosion of a
burst. It is natural that this value is larger than others which
are measured at later times. This might be the cause for the
detected difference. Ignoring this slight difference, our method
is consistent with that referring to the delayed emission of the
early afterglow.

We suspect that, a very strong shock might produce higher energy
photons, which is characterized by a large value of $E_{0,p}$, and
this would lead to a large value of $E_p$ (note that, as shown
above, the Lorentz factor does
not change much for different sources). We make a statistical analysis for $%
E_\gamma $ and $E_{0,p}$ and find that they are indeed obviously
correlated (the figure is omitted). We then understand why
$E_\gamma $ is correlated
with $E_p$. It is because that strong shocks produce large values of both $%
E_\gamma $ and $E_p$, whereas weak shocks lead to smaller values.

We assume through out this paper that the afterglow is dominated
by an adiabatic process. However, there is an alternative which is
a radiative process, for which, $p=3/7$ should be adopted. We
repeat our work by replacing $p=3/8$ with $p=3/7$. We find that,
the value of $\Gamma $ is mainly distributed within $(106,584)$
which is slightly larger than what we expect. Thus, we tend to
believe that, during the epoch of the afterglow, the dominated
process is adiabatic rather than radiative (this is not certain
since the resulted Lorentz factors are still within the acceptable
range).

In the above analysis, we assume that $\xi (\Gamma
-1)mc^{2}\gg\Delta E_{th}$. Does our analysis strongly depend on
this assumption? To find an answer to this, let us assign $\Delta
E_{th} \equiv \eta \xi (\Gamma -1)mc^{2}$, where $\eta$ is a
constant. In this way, equation (4) becomes
\begin{equation}
\Gamma mc^{2}-\Gamma _{dur}mc^{2} =(1+ \eta) \xi (\Gamma
-1)mc^{2}.
\end{equation}
Taking $\xi'\equiv(1+ \eta) \xi$, we get
\begin{equation}
(1-\xi' )\Gamma  =\Gamma _{dur}-\xi'.
\end{equation}
According to (10), $\xi'=1$ suggests that all the explosion energy
converts to radiation, which would not be true. Thus, we have
$0<\xi'<1$. In this case, one finds that formulas (5), (7), (8)
and (9) are valid when one replaces $\xi$ with $\xi'$. This
indicates that the above analysis does not depend on the the
mentioned assumption. When the two amounts of energy are
comparable, i.e., $ \eta\simeq 1$, one has $\xi'\simeq 2 \xi$.
Adopting $\xi=0.2$ leads to $\xi'=0.4$, which would not
significantly change the results obtained above.
\section*{Acknowledgments}

This work was supported by the Special Funds for Major State Basic
Research Projects (``973'') and National Natural Science
Foundation of China (No. 10273019).

%====================

\end{document}